\documentclass[12pt]{iopart}

\usepackage{epsfig}
\usepackage{graphicx}
\usepackage[dvips]{color}
\usepackage{color}

\begin{document}

\title[Numerical studies of the electron-positron filamentation 
instability in 1D]{Simulation study of the filamentation of 
counter-streaming beams of the electrons and positrons in plasmas}

\author{M E Dieckmann$^{1,2}$, P K Shukla$^{2,3}$ and L Stenflo$^{4}$}

\address{1 Department of Science and Technology, Link\"oping University, 
SE-60174 Norrk\"oping, Sweden \\
2 Institute of Theoretical Physics IV, Ruhr-University Bochum, 
D-44780 Bochum, Germany \\
3 SUPA Department of Physics, University of Strathclyde, Glasgow,
Scotland, and Department of Physics, Ume\aa \, University, SE-90187
Ume\aa, Sweden \\
4 Department of Physics, Chemistry and Biology, Link\"oping University,
SE-58183 Link\"oping, Sweden}
\ead{markd@tp4.rub.de}

\begin{abstract}
The filamentation instability driven by two spatially uniform and
counter-streaming beams of charged particles in plasmas is modelled by
a particle-in-cell (PIC) simulation. Each beam consists of the electrons 
and positrons. The four species are equally dense and they have the same 
temperature. The one-dimensional simulation direction is orthogonal 
to the beam velocity vector. The magnetic field grows spontaneously and 
rearranges the particles in space, such that the distributions of the 
electrons of one beam and the positrons of the second beam match. The 
simulation demonstrates that as a result no electrostatic field is 
generated by the magnetic field through its magnetic pressure gradient
prior to its saturation. This electrostatic field would be repulsive 
at the centres of the filaments and limit the maximum charge and current 
density. The filaments of electrons and positrons in this simulation 
reach higher charge and current densities than in one with no positrons. 
The oscillations of the magnetic field strength induced by the magnetically 
trapped particles result in an oscillatory magnetic pressure gradient force. 
The latter interplays with the statistical fluctuations in the particle 
density and it probably enforces a charge separation, by which electrostatic 
waves grow after the filamentation instability has saturated.
\end{abstract}


\maketitle

\section{Introduction}

If a plasma is initially free of any net current, but if it has a 
nonequilibrium particle velocity distribution, then it can support the
growth of magnetic fields through the redistribution of currents in space. 
This has been demonstrated first by Weibel \cite{Weibel} for a plasma with 
a bi-Maxwellian electron velocity distribution. The currents are rearranged 
through the growing plasma waves into filaments, which are separated by 
electromagnetic fields \cite{Morse}. This is also the case for the 
instability driven by counterstreaming beams of charged particles, which 
is commonly referred to as the beam-Weibel instability or the filamentation 
instability (FI) \cite{Lee,Gremillet,Bret,Bret2}. 

The FI can generate strong magnetic fields in astrophysical environments 
such as the leptonic pulsar winds \cite{Arons}. The FIs are also important 
for the generation of cosmological magnetic fields \cite{Schlickeiser} and 
for inertial confinement fusion \cite{Wilks,Martin}, where laser pulses 
accelerate electron beams to relativistic speeds. Previous simulation 
studies have revealed various aspects of the growth and saturation of the 
FI. The FI driven by two counter-propagating beams of electrons has been 
examined by using particle-in-cell (PIC) and Vlasov simulations. Such 
studies have been performed in one spatial dimension (1D) 
\cite{Davidson,Cal1,Cal2,Rowlands,Dieckmann2}, in two spatial 
dimensions (2D) \cite{Lee,Medvedev,Dieckmann,Stockem} and in 3D \cite{Sakai}. 
The effects of a guiding magnetic field on counter-streaming electron 
beams have also been examined \cite{Stockem,BreDi}. The simulation studies 
in Refs. \cite{Cal2,Pukhov} have investigated the impact of the 
ions on the non-linear stage of the FI.

The FIs are usually triggered by the electrons. The ion filamentation is 
slower and often coupled through electrostatic fields to the electron
filamentation. The FI involving only electrons couples to electrostatic 
fields during its quasi-linear growth phase. We refer to this source 
mechanism of electrostatic waves as the quasi-linear electrostatic 
instability (QEI). The electrostatic field amplitude grows in response
to the QEI at twice the exponential rate of the magnetic field amplitude 
\cite{Cal1,Cal2,Rowlands} and it oscillates around an equilibrium value 
after the FI has saturated. This equilibrium amplitude is such, that it 
excerts a force on the electrons that equals that of the average magnetic 
pressure gradient force (MPGF) \cite{Dieckmann2}. This was demonstrated 
for the case of two counter-streaming, equally dense electron beams, for 
which the growth rate of the purely transverse FI is highest relative to 
those of the competing electrostatic two-stream instability and of the 
partially electromagnetic mixed mode instability \cite{Bellido}. The 
omission of wavevectors aligned with the beam velocity vector is thus 
most realistic. 

The pulsar winds carry with them positrons \cite{Arons}. The impact of the 
positrons on the initial growth phase of the filamentation instabilities 
is understood, in principle, by solving the linear dispersion relation, as 
it has been done for example in Refs. \cite{Fiore,Tautz,Petri2} for the FI 
and in Ref. \cite{Petri} for the Weibel instability. The nonlinear evolution
of instabilities driven by beams of the electrons and positrons have been 
modelled with PIC simulations in 2D \cite{Tautz2,Kazimura} and in 3D 
\cite{Silva}. 

In this paper, we consider two counterstreaming plasma beams, each of which 
contains an identical number of positrons and electrons. The FI is modelled 
in a short 1D simulation box, in which only one filament pair develops, as 
it has been done in Refs. \cite{Cal1,Cal2,Dieckmann2} for the electron beams. 
The non-linear saturation of the FI is not captured correctly, since we 
exclude the filament merging and the multi-dimensional structure of the 
filaments \cite{Lee}. However, the filaments are not circular if the
beams are warm and if they have the same density. We find spatial intervals
of the filaments that are planar over several electron skin depths 
\cite{Dieckmann,Stockem,SilvaAIP}. A 1D simulation can give insight into
the dynamics of such planar boundaries. The 1D geometry allows us to freeze 
the filament pair just after the initial saturation and we can analyse the 
filaments in an almost time-stationary form. We can isolate a single 
filament pair to better understand its dynamics by omitting its collective 
interactions with the neighboring filaments. The restriction to one spatial 
dimension furthermore permits us to use a good statistical representation 
of the plasma phase space distribution and we can reduce the simulation 
noise. Accurate measurements of the fields and of the phase space distribution 
are thus possible. We choose initial conditions, which are identical to those 
in Refs. \cite{Rowlands,Dieckmann2}, except for the positronic beam component 
that we include here. This allows us to compare directly the electromagnetic 
interaction of two counterpropagating electron beams with that of two 
counterpropagating beams of the electrons and positrons. 

We summarize our key results. The symmetry between the electrons and the
positrons suppresses the QEI during the quasi-linear growth phase of the 
FI. The magnetic trapping model \cite{Davidson}, which does not consider 
an electric field, accurately describes the saturation magnetic field
in the simulation. The electrostatic 
field driven by the QEI would repel the electrons at the centre of the 
filament and attract those farther away \cite{Dieckmann2}, thus limiting 
the charge density accumulation due to the FI. Its absence implies that 
the filament confinement is stronger, if the positrons are present; higher 
peak density values can be reached and the spacing between the filaments 
is larger. The magnetic fields should reach higher spatial gradients 
compared to a system of counter-propagating electron beams due to the 
stronger currents. We could separate the purely magnetic FI driven by 
counter-propagating electron-positron beams from a secondary electrostatic 
instability (SEI), which we show to be unrelated to the QEI. The waves 
the SEI drives have a broadband wavenumber spectrum. We bring forward 
evidence for a connection between the SEI and the spatio-temporal 
oscillations of the MPGF. These oscillations occur on time scales, which 
are comparable to the inverse plasma frequency, and on spatial scales of
the order of a Debye length. The MPGF can thus interplay through these 
oscillations with the statistical fluctuations of the plasma, by which a 
charge separation can occur. We propose that this separation breaks the 
initial symmetry of electrons and positrons and destabilizes the filament. 
The power of the waves driven by the SEI, which have a frequency that is 
close to the plasma frequency, grows in time. Its power can be fitted as a 
function of time by two exponential functions, which are separated by a 
break in the growth rate. 

The manuscript is organized as follows. Section 2 briefly describes the PIC 
simulation method and it discusses our initial conditions. Section 3 
presents our simulation results, which are then discussed in Section 4.

\section{The PIC simulation method and the initial conditions}

The standard particle-in-cell (PIC) method \cite{Dawson} can model the 
processes in a collisionless kinetic plasma. It approximates the plasma 
phase space distribution by an ensemble of volume elements or computational 
particles (CPs). The ensemble properties of the CPs are an approximation
to the ensemble properties of the corresponding physical plasma species. 
Each CP with index $i$ of the species $j$ can have a charge $q_j$ and mass 
$m_j$ that differ from those of the plasma species they represent, e.g. the 
mass $m_e$ and the charge $-e$ of an electron. The charge-to-mass ratio must 
be preserved though. 

The CPs follow trajectories in the simulation domain, which are determined 
by the electric $\mathbf{E}$ and the magnetic $\mathbf{B}$ fields. The 
electromagnetic fields and the global current $\mathbf{J}$ are defined on 
a spatial grid. The equations that are solved by the PIC code are 
\begin{eqnarray}
\nabla \times \mathbf{E} = -\frac{\partial \mathbf{B}}{\partial t}, \, \, \,
\, \, \nabla \times \mathbf{B} = \frac{\partial \mathbf{E}}{\partial t} + 
\mathbf{J}, \label{eq1} \\
\nabla \cdot \mathbf{B} = 0, \, \, \, \, \, \nabla \cdot \mathbf{E} = \rho, 
\label{eq2} \\
\frac{d\mathbf{p}_{i}}{dt} = q_j \left ( \mathbf{E}[\mathbf{x}_i] + 
\mathbf{v}_i \times \mathbf{B}[\mathbf{x}_i] \right ), \, \, \, \, \, 
\mathbf{p}_{i} = m_j \, \mathbf{v}_i \, \Gamma (\mathbf{v}_i), \, \, \, \, \, 
\frac{d\mathbf{x}_i}{dt} = \mathbf{v}_{i}. \label{eq3}
\end{eqnarray}
The electromagnetic fields are evolved in time using the Faraday law and 
the Ampere law (Eq. (\ref{eq1})). Equations (\ref{eq2}) are typically 
fullfilled as constraints, or they are enforced by correction steps. The 
virtual particle method \cite{Eastwood} our code is using is fullfilling 
Poisson's equation as a constraint and $\nabla \cdot \mathbf{B} = 0$ is 
solved exactly in 1D and to round-off precision in higher dimensions. The 
trajectories of the CPs are updated using the Eqns. (\ref{eq3}). We refer 
the interested reader to the Refs. \cite{Dawson,Eastwood,S1,S2,S3,S4} 
for a more thorough discussion of numerical PIC schemes.

Equations (1)-(3) can be scaled to physical units with the total plasma 
frequency $\omega_p = {(n_t e^2 / m_e \epsilon_0)}^{1/2}$ and the skin depth 
$\lambda_e = c / \omega_p$, where the total particle number density 
$n_t = \sum_j n_j$ is summed over the four leptonic species, which are 
equally dense. The quantities in physical units denoted by the subscript 
$p$ are obtained from the normalized ones by substituting $\mathbf{E}_p = 
\omega_p c m_e \mathbf{E} / e$, $\mathbf{B}_p = \omega_p m_e \mathbf{B} / 
e$, $\mathbf{J_p} =e c n_t \mathbf{J}$, $\rho_p = e n_t \rho$, $\mathbf{x}_p 
=\lambda_e \mathbf{x}$, $t_p = t / \omega_p$, $\mathbf{v}_p = \mathbf{v}c$ 
and $\mathbf{p}_p = m_e c \mathbf{p}$. The charge $q_j$, in this 
normalization, is 1 for the positrons and -1 for the electrons, while 
$m_j = 1$. We also normalize $\Omega = \omega / \omega_p$ and $\mathbf{k} 
= \mathbf{k}_p c / \omega_p$, where $\omega,\mathbf{k}_p$ have physical units. 

Both beams in our simulation study move in opposite $\mathbf{z}$-directions 
with the speed modulus $v_b = 0.3$, giving the relative beam speed $2v_b / 
(1+v_b^2) \approx 0.55$. This $v_b$ is about half that used in Ref. 
\cite{Kazimura}. Each of the two beams consists of one electron species 
and of one positron species and all four species are equally dense. Initially 
all beams are spatially uniform. The velocity distribution in the rest frame 
of each beam is a Maxwellian with the thermal speed $v_t \equiv 
{(kT/m_e)}^{1/2} = v_b / 18$ in all directions. The 1D simulation box with 
its periodic boundary conditions is aligned with the $\mathbf{x}$-direction. 
Only waves with wavevectors parallel to $x$ can grow and we use the scalar 
wavenumber $k$. The length $L=1.25 \lambda_e$ of the simulation box is 
identical to that of the shortest one in Ref. \cite{Dieckmann2}, if we 
neglect the positron contribution to $\lambda_e$. The simulation box is 
subdivided into 500 grid cells of equal length $\Delta_x$. The ratio
$\Delta_x \omega_p / v_t = 0.15$ and the Debye length is resolved well.
Each of the four plasma species is represented by $2.45 \times 10^7$ CPs. 
The $\mathbf{E}=0$ and $\mathbf{B}=0$ at the simulation's start. The total 
simulation time is $t_{sim}=177$, which is subdivided into $10^5$ time steps 
$\Delta_t$. 

\section{Simulation results}

The FI driven by a plasma flow along $z$ and a simulation box that is 
aligned with $x$ will lead to the initial growth of a magnetic field 
along $y$. The growing net current $J_z (x)$ will also result in a
growing $E_z$ by Ampere's law. An electrostatic $E_x$ field would 
grow in the case of a system that is composed of two electron beams. 

\begin{figure}
\centering
\includegraphics[width=8.2cm]{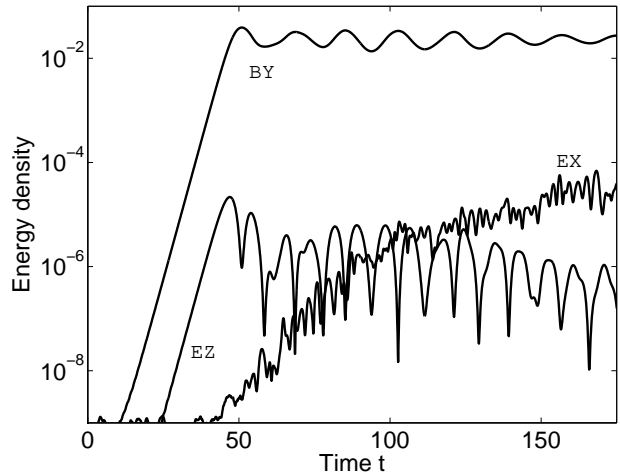}
\caption{The energy densities in units of the total energy in the 
simulation: The uppermost curve $E_{BY}$ corresponds to $B_y$. Its 
exponential rate, which is twice the $\Omega_i$ of $B_y$, is $\approx 
0.5$ during $10<t<50$. It oscillates around an equilibrium after that. 
The curves of $E_{EZ}$ and $E_{BY}$ grow initially at the same exponential 
rate, but $E_{EZ}$ decreases for $t>50$. The electrostatic $E_{EX}$ starts 
to grow at $t\approx 45$, when $E_{BY}$ has saturated.}\label{Plot1}
\end{figure}

Figure \ref{Plot1} displays the energy densities of these field components, 
which we denote as $E_{BY}$, $E_{EZ}$ and $E_{EX}$.  
The energy density $E_{BY}$ of the magnetic $B_y$ component dominates and 
it reaches a few percent of the total energy, in line with previous
simulations \cite{Kazimura}. The exponential growth rate of $B_y$ is 
$\Omega_i \approx 0.25$, which is close to the expected value $\approx 
v_b / \gamma^{1/2}_b$ for cold beams. The $E_{EZ}$ grows at the same rate 
as $E_{BY}$, but its values are three orders of magnitude less. The energy 
density $E_{EX}$ of the electrostatic field grows after the FI has saturated. 
The growth rate of $E_{Ex} \propto E_x^2$ can be fitted with an exponential 
function with the growth rate $\Omega_{ix} \approx 0.13$ between $45<t<90$ 
and with a second, slower growing one for $t>100$. The growth of $E_{EX}$ 
in Fig. \ref{Plot1} can not be attributed to the QEI, because then the 
$E_{EX}$ should grow at twice the exponential rate of $E_{BY}$ until 
$t \approx 50$ and oscillate around an equilibrium value after that time.

\begin{figure}
\centering
\includegraphics[width=7.5cm]{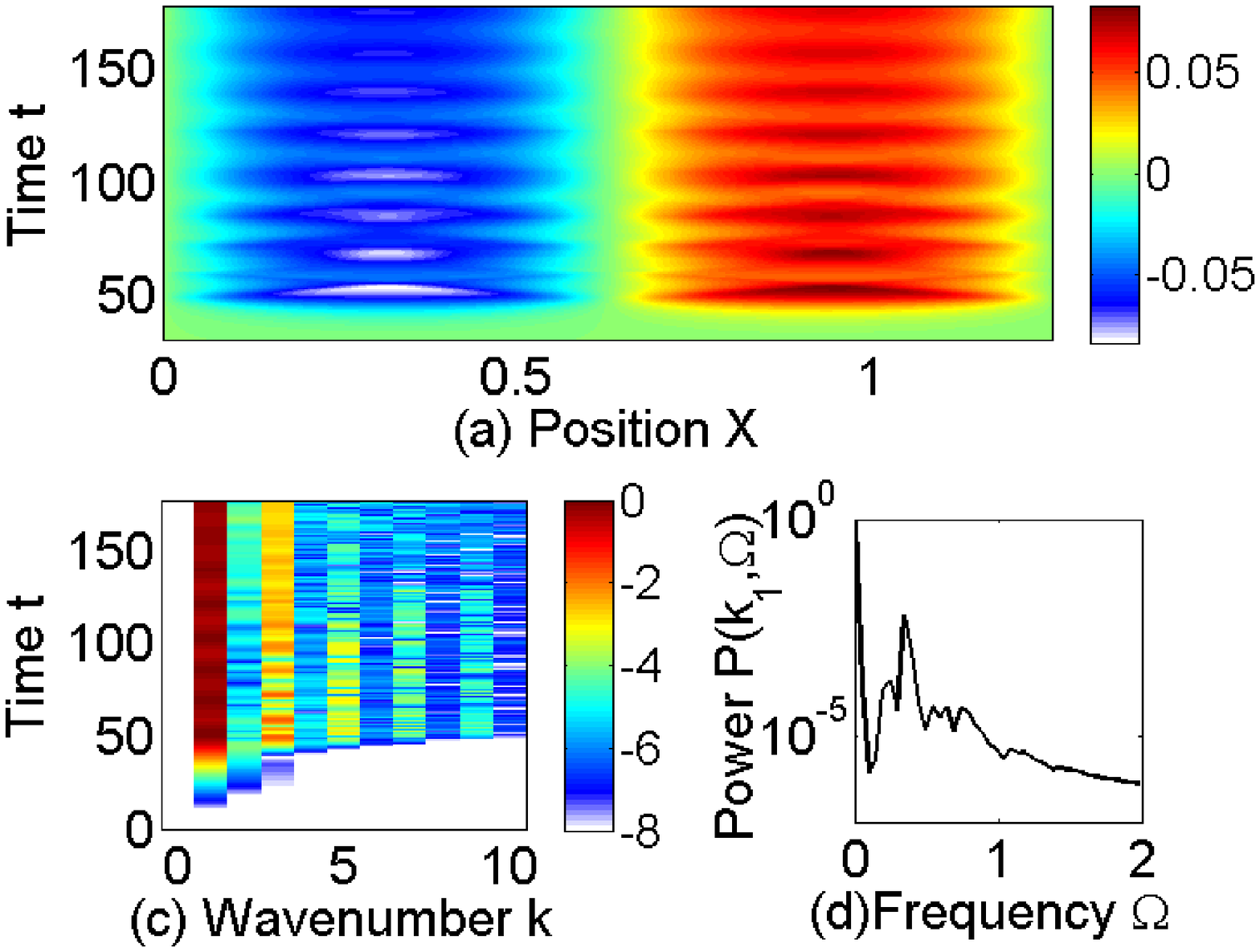}
\includegraphics[width=7.5cm]{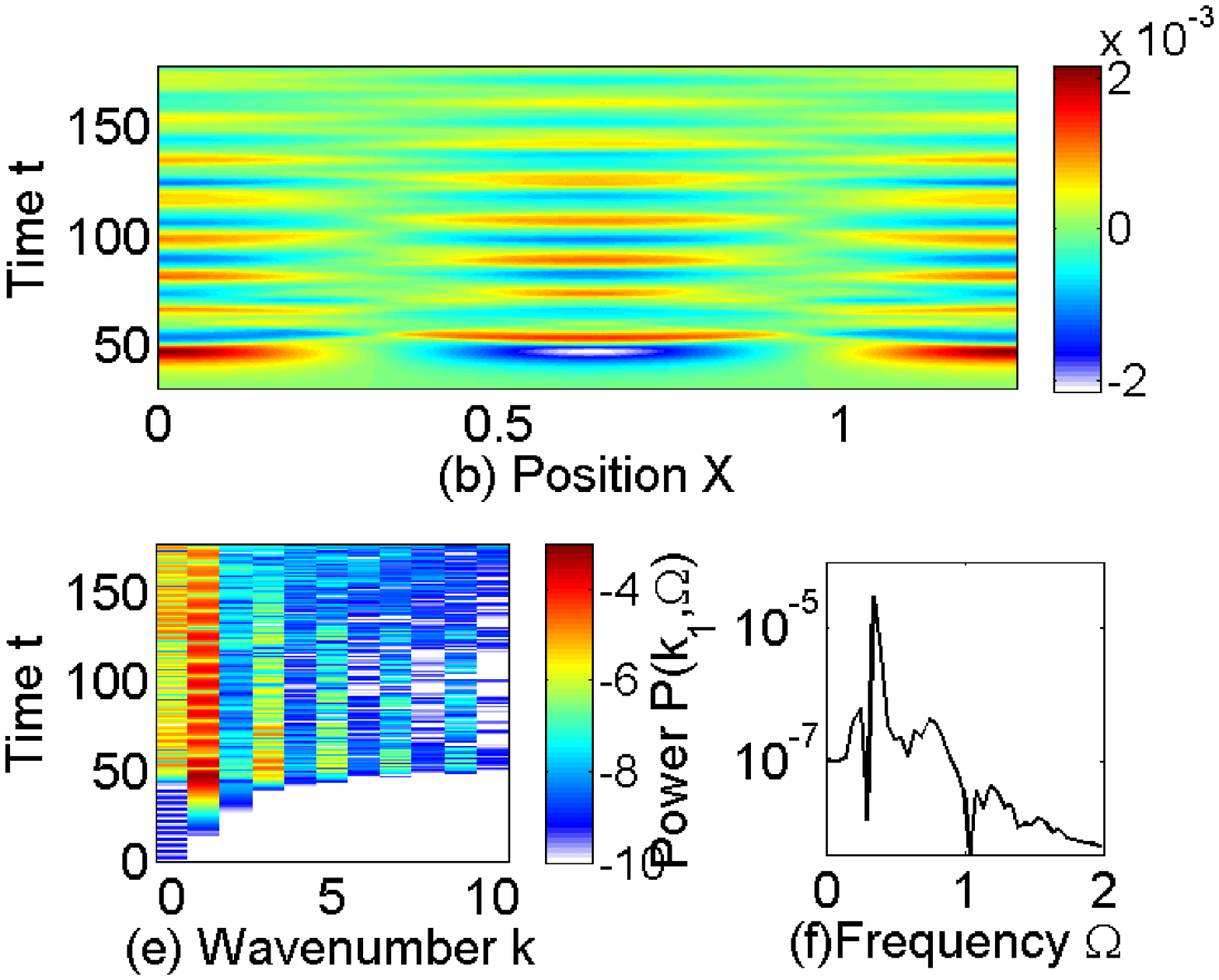}
\caption{(Colour online) The electromagnetic fields: Panels (a) and (b) show 
$B_y$ and $E_z$, respectively. Both are correlated and their phase difference 
is $90^\circ$. The $E_z$ field is damped in time. Panel (c) shows the spatial 
power spectrum $log_{10} P(k,t)$ of $B_y$. The $k_1$ mode dominates, but 
harmonics can be seen. Panel (d) shows the frequency power spectrum 
$log_{10} P(k_1,\Omega)$ of $B_y$. Peaks are found at $\Omega =0$ and 
$\Omega \approx 0.3$. Panel (e) evidences a $log_{10}P(k,t)$ of $E_z$ 
resembling that of $B_y$ in (c) and the $\log_{10}P(k_1,\Omega)$ of $E_z$ 
in (f) also has a maximum at $\Omega \approx 0.3$. The power in (c,e) and 
(d,f) is normalized to the same value.}\label{Plot2}
\end{figure}

We look in more detail at the fields to better understand the cause of the 
oscillations of $E_{BY}$ around its equilibrium and the source mechanism 
of the growth of $E_{EX}$. The field components driven by the FI are
investigated in Fig. \ref{Plot2}. The $B_y, E_z$ saturate at $t\approx 50$. 
The $B_y(x,t)$ then remains practically stationary, while $E_z(x,t)$ is 
damped. 

We compute the spatial power spectrum $P(k,t)$ of $B_y,E_z$ and $E_x$ by a 
1D Fourier transform over the full box length $L$ and by taking the square 
of the amplitude modulus. The frequency power spectra (dispersion relation) 
are obtained through a 2D Fourier transform of the field data over the box 
length $L$ and for $t>50$. The amplitude moduli are then squared to give 
$P(k,\Omega)$. The base-10 logarithm of $P(k,t)$ of $B_y$ in Fig. 
\ref{Plot2}(c) evidences, that most power is concentrated in the mode $k_1$, 
with $k_s = 2\pi s / L$, and that the power in this mode is oscillating. 
Weaker harmonics with uneven $s$ also occur. The $P(k_1,\Omega)$ of $B_y$ 
reveals peaks at $\Omega = 0$ and $\Omega \approx 0.3$. The peak at $\Omega 
= 0$ dominates, because $B_y$ is practically stationary after $t=50$. The 
$P(k,t)$ and the $P(k_1,\Omega)$ of $E_z$ in the Figs. \ref{Plot2}(e,f)
resemble qualitatively those of $B_y$, but they are weaker and the peak 
with $\Omega = 0$ is absent in the $E_z$-field.

\begin{figure}
\centering
\includegraphics[width=8.2cm]{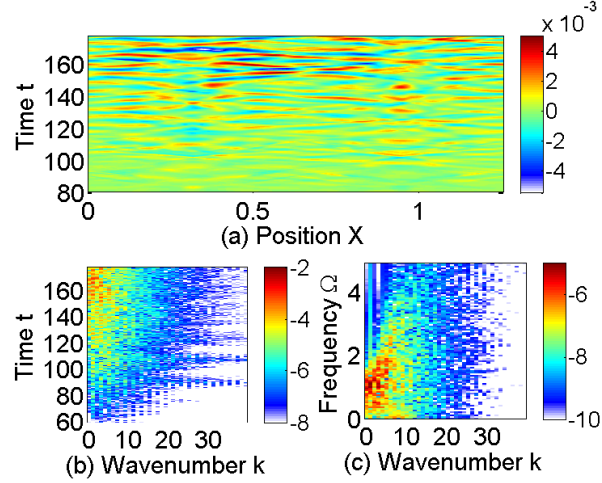}
\caption{(Colour online) The electrostatic field: Panel (a) shows $E_x(x,t)$. 
Fluctuations grow at late times, which are distributed seemingly randomly 
in space. Its $\log_{10}P(k,t)$ in (b) evidences a broadband spectrum, that 
shows no dominant $k$. The $\log_{10} P(k,\Omega)$ peaks at $\Omega \approx 
1$ and $k>0$. The strongest fluctuations are propagating waves. The $P(k,t)$
and $P(k,\Omega)$ are normalized to the corresponding values of $B_y$.}
\label{Plot3}
\end{figure}
The $E_x (x,t)$ in Fig. \ref{Plot3} reveals no spatial correlation 
with $B_y (x,t)$. The $P(k,t)$ of the $E_x$-field in Fig. 
\ref{Plot3}(b) and of the $B_y$-field in Fig. \ref{Plot2}(c) show no 
link between the $k$ of the dominant waves. The spatial power spectrum 
$P(k,t)$ in Fig. \ref{Plot3}(b) evidences instead a wave growth over a 
wide band of $k$. The $E_x$ can not be driven by the QEI, because then 
the MPGF, which is $\partial_x \tilde{P}_{BY} = B_y dB_y / dx$ in our 
normalization, would imply that $E_x = 0$ whenever $B_y = 0$ or $dB_y / 
dx = 0$ and that it should oscillate in the $k_2$ mode. The 
$P(k,\Omega)$ of $E_x$ in Fig. \ref{Plot3}(c) reveals that the strongest 
waves have a $\Omega \approx 1$.

\begin{figure}
\centering
\includegraphics[width=5cm]{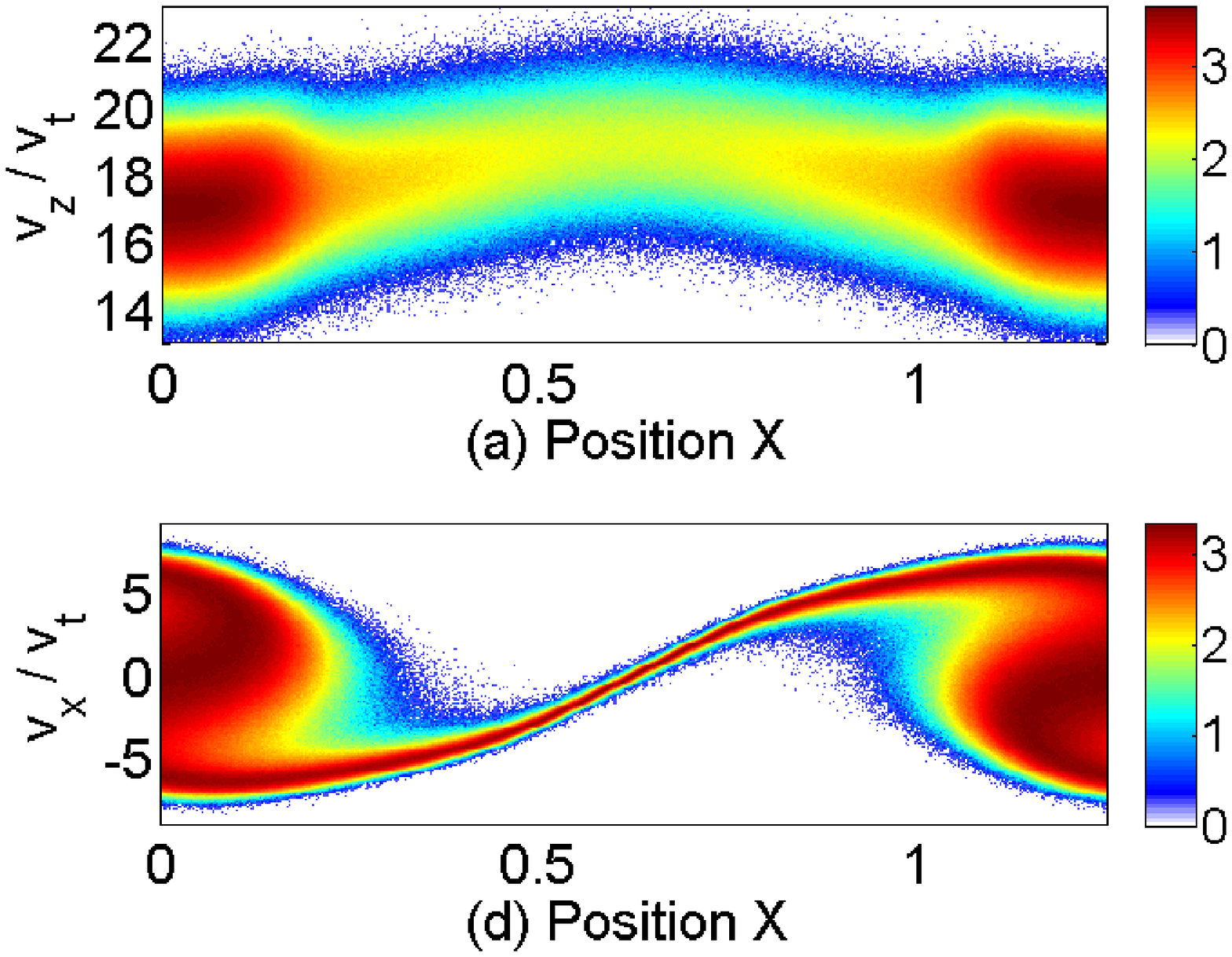}
\includegraphics[width=5cm]{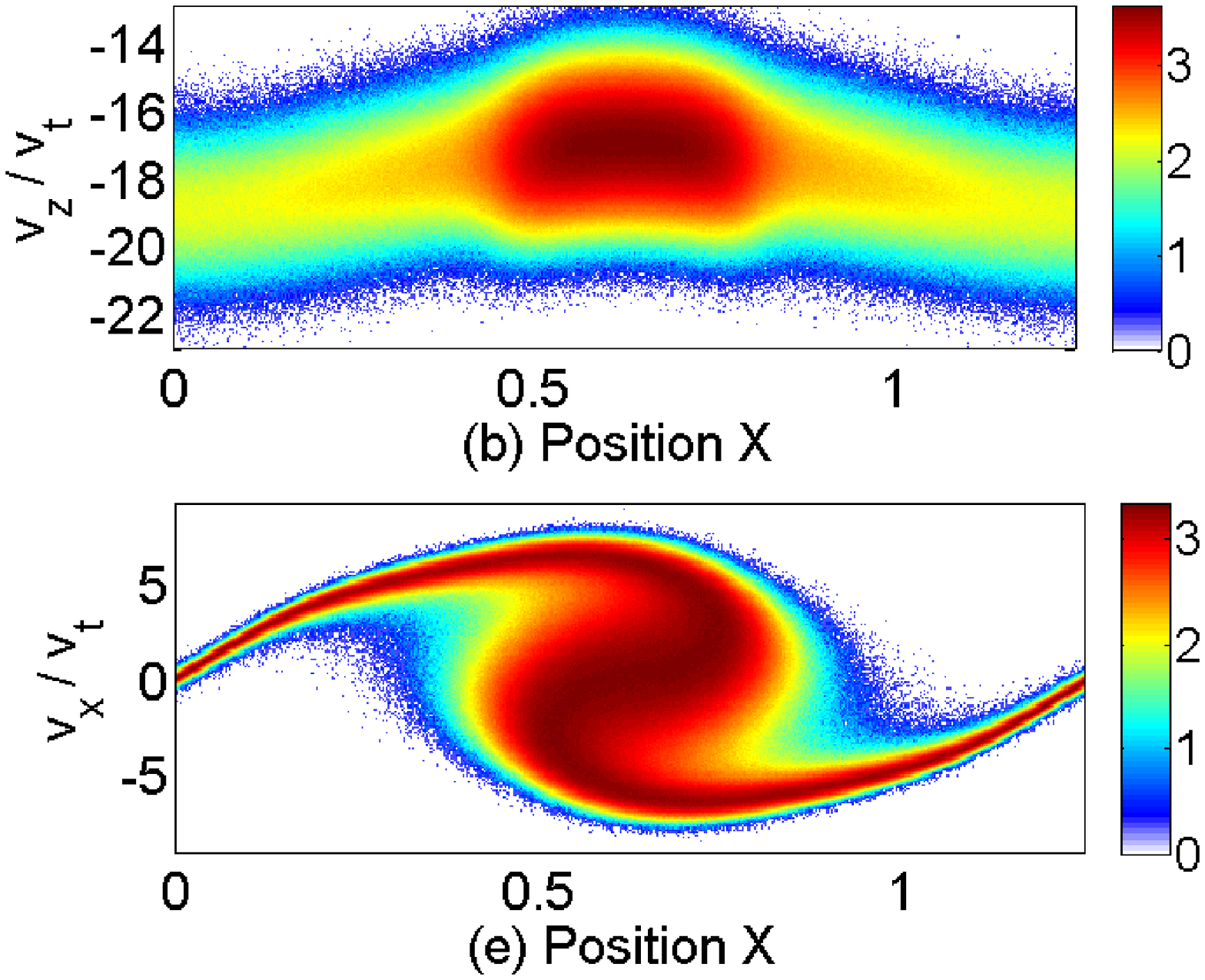}
\includegraphics[width=5cm]{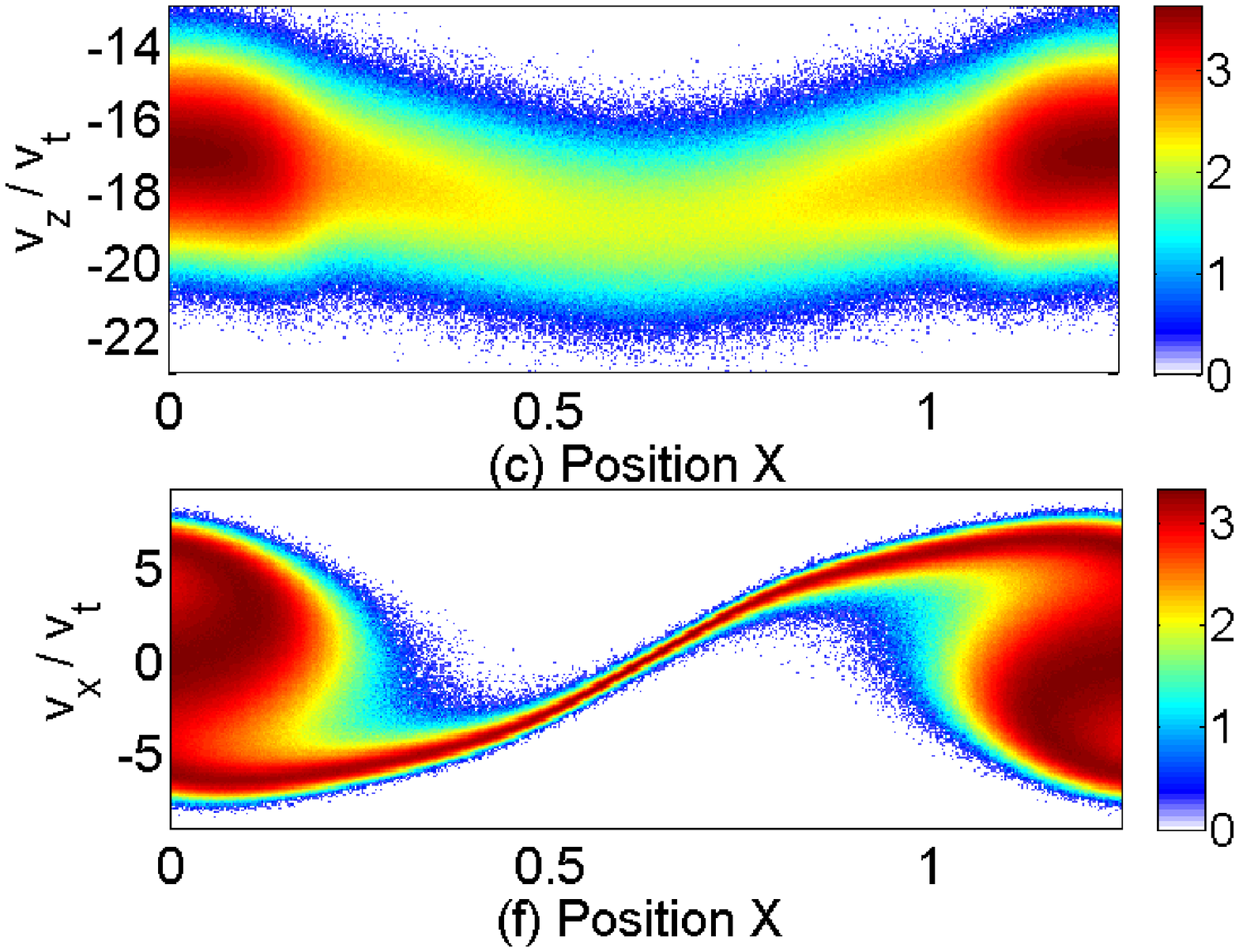}
\caption{(Colour online) The base-10 logarithm of the phase space 
densities at the time $t=50$ in units of CPs: Panels (a-c) display the 
$f(x,v_z)$ of the species 1, 2 and 4 (positrons). (d-f) show the $f(x,v_x)$ 
corresponding to the panels above them. The distributions in (d,f) are 
practically identical and (a,c) can be mapped onto each other by changing 
the sign of $v_z$. The $f(x,v_x)$ of species 1 and 2 are shifted by L/2.}
\label{Plot4}
\end{figure}

The particle phase space distributions provide more information. The 
electrons of the beam 1 (moves in positive z-direction) are species 1 
and the positrons are species 3. The electrons and positrons of beam 2 
are denoted as species 2 and 4, respectively. The current fluctuations of 
PIC simulations imply that $B_z,E_y \neq 0$. These fields correspond to 
waves that propagate in form of the high-frequency electromagnetic modes. 
However, the peak value of $E_{EZ}$ exceeds the energy densities of 
$E_y,B_z$ by 6 orders of magnitude and the latter remain practically 
constant during the simulation time. We can restrict our investigation 
to the phase space projections $f(x,v_x)$ and $f(x,v_z)$, because $B_x = 0$, 
$B_z \approx 0$ and $E_y \approx 0$. The $f(x,v_z)$ will reveal the 
electromagnetic structures, while the electrostatic structures are 
represented by $f(x,v_x)$. Figure \ref{Plot4} shows the phase space 
distributions of the species 1,2 and 4 at $t=50$. 
The electrons of beam 1 and 2 are separated in space, which is typical
of the FI driven by counter-streaming electron beams. The species 4
shows an $f(x,v_x)$ that is identical to that of species 1. The $f(x,v_z)$ 
distribution of species 1 can be mapped onto that one of species 4 by 
switching the sign of $v_z$. Both these observations are expected, 
because in the absence of a significant $E_x$ the opposite speed and 
charge of both species cancel. The Lorentz force thus displaces both 
species in the same way. The relation is also the same between the species 
2 and 3. The high degree of symmetry is also demonstrated by the 
movie 1, which animates in time the projected phase space distributions 
of species 1, which is added to that of species 4. The color scale shows 
the base-10 logarithm of the number of CPs.

\begin{figure}
\centering
\includegraphics[width=8.2cm]{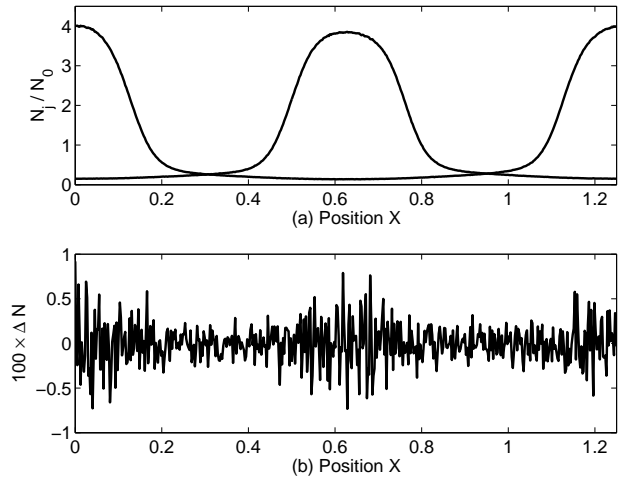}
\caption{Panel (a) The number density distribution $N_j(x,t=50)$ of the four 
species $j$ in units of their mean density $N_0$. The curves $N_1(x)$ and 
$N_4(x)$ match (their differences amount to less than the thickness of the
curves) and they peak at $x=0$. The also matching $N_2 (x)$ and $N_3(x)$ have 
their maximum at $x=L/2$. Panel (b) plots $\Delta N = \sum_j q_j N_j(x) / 
4N_0$, with $q_j=1$ and $-1$ being the positron and electron charge. The 
fluctuations amount to less than $10^{-2}$ and they are by their scale size 
$\approx \Delta_x$ the statistical fluctuations due to the finite number 
of CPs per cell.}\label{Plot5}
\end{figure}

Figure \ref{Plot5} displays the number densities $N_i (x) = \int f_i(x,v_x) 
dv_x$ for each of the species $i$, normalized to the mean value $N_0 = 
\left < N_i (x) \right >_x$.
The $N_1 = N_4$ and $N_2 = N_3$ within the resolution of the image. The 
charge density modulations reach a peak value $\approx 4 N_0$. This peak
value is higher by a factor three and, accordingly, the filament confinement 
is thus stronger here than in Ref. \cite{Dieckmann2}, which did not consider 
positrons. Each species is represented by $N_p = 4.9 \times 10^4$ CPs per 
cell. The statistical fluctuations of the particle number are thus 
${N_p}^{-0.5} \approx 5 \times 10^{-3}$, which is comparable to the measured 
charge density fluctuations $\Delta N = {(4N_0)}^{-1} \sum_i q_i N_i < 10^{-2}$ 
in Fig. \ref{Plot5}(b). This observation, together with that the $\Delta N$ 
oscillates on a scale $\Delta x$, implies that the fluctuations at $t=50$ 
are due to the finite number of CPs per cell. These fluctuations have a 
$k$-spectrum that is qualitatively similar to that of thermal noise. The
fluctuation amplitude $\Delta N$ is increased within the filaments, e.g.
in the interval $0.5 < x < 0.75$, because we have not normalized it to 
the local particle number, but to the average density.

\begin{figure}
\centering
\includegraphics[width=5cm]{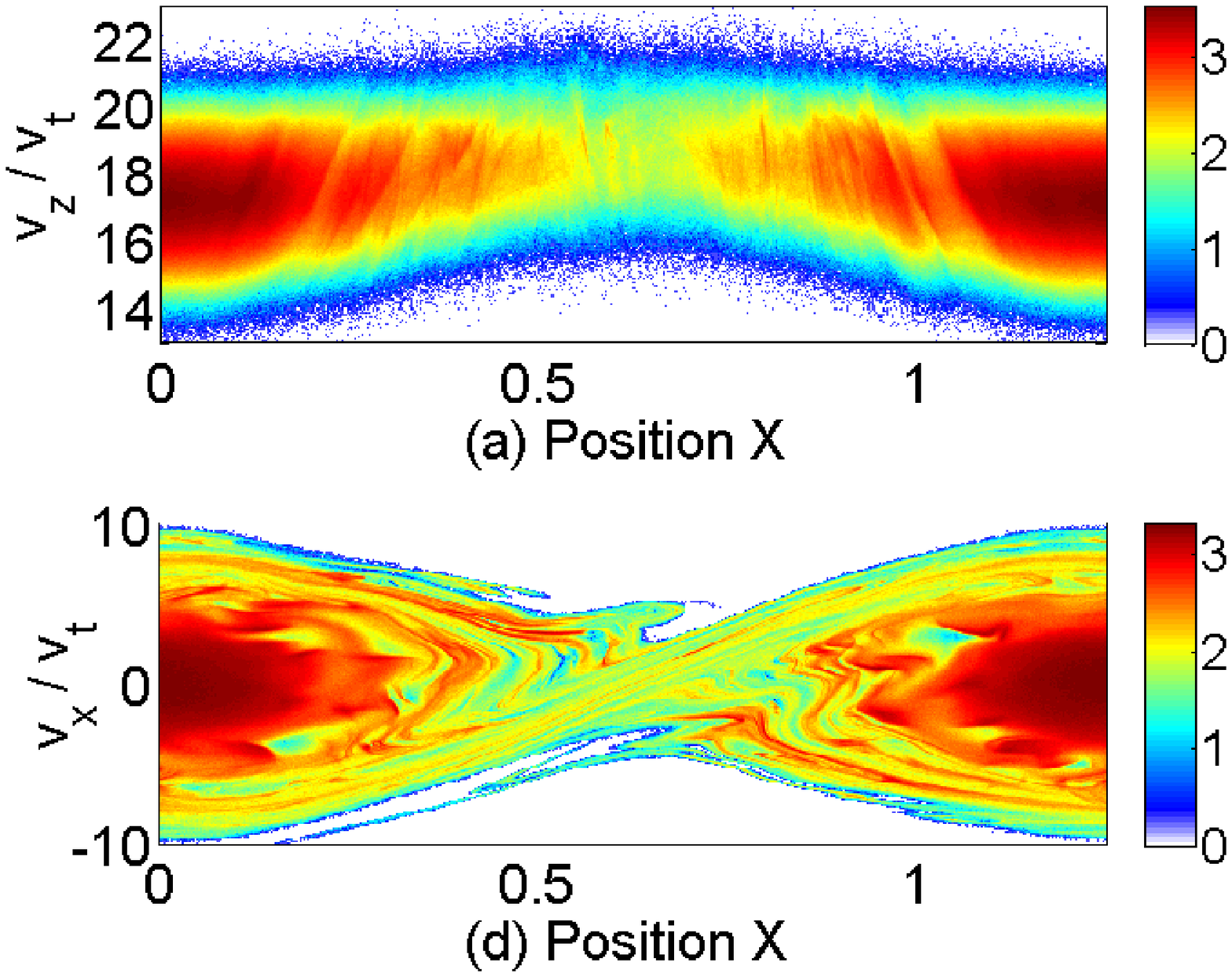}
\includegraphics[width=5cm]{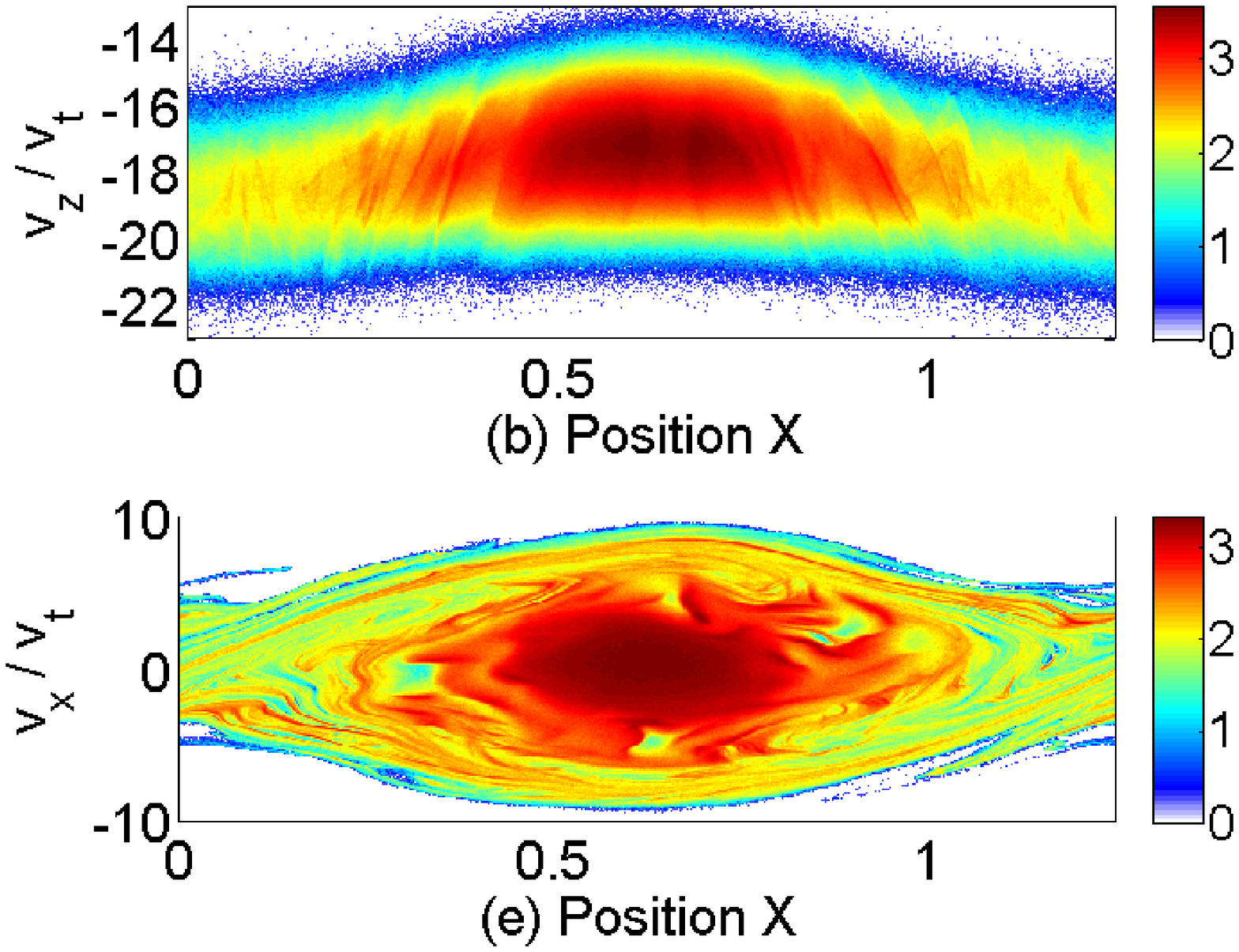}
\includegraphics[width=5cm]{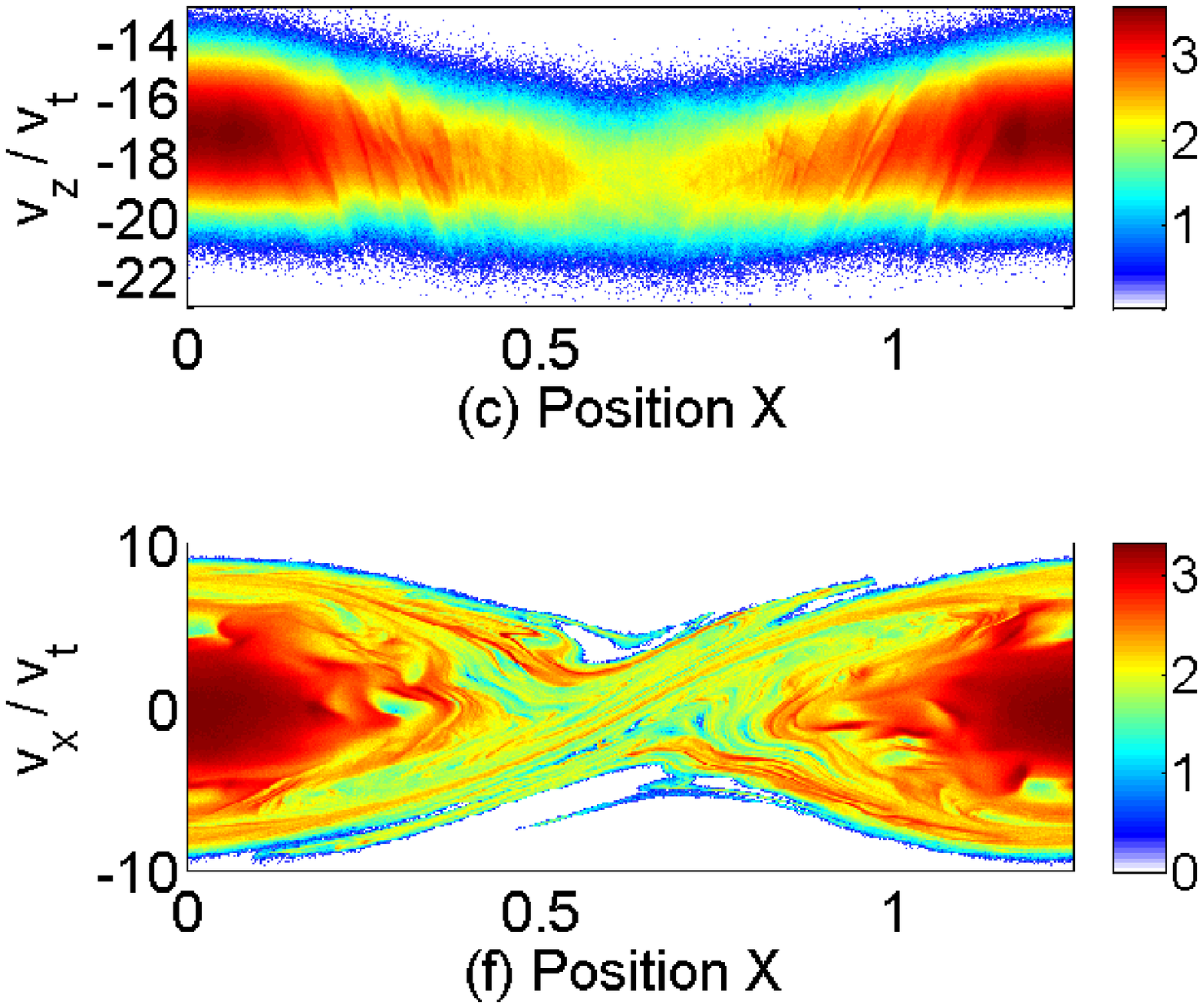}
\caption{(Colour online) The base-10 logarithm of the phase space densities 
at the time $t=177$ in units of CPs: Panels (a-c) show $f(x,v_z)$ of the 
species 1, 2 and 4, respectively. (d-f) show the $f(x,v_x)$ corresponding to 
the panels above. Species 1 and 4 reveal similar $f(x,v_x)$ and distributions 
$f(x,v_z)$ that are qualitatively symmetric to a change of the sign of $v_z$. 
The $f(x,v_x)$ of species 1 and 2 are shifted by L/2.}\label{Plot6}
\end{figure}

\begin{figure}
\centering
\includegraphics[width=8.2cm]{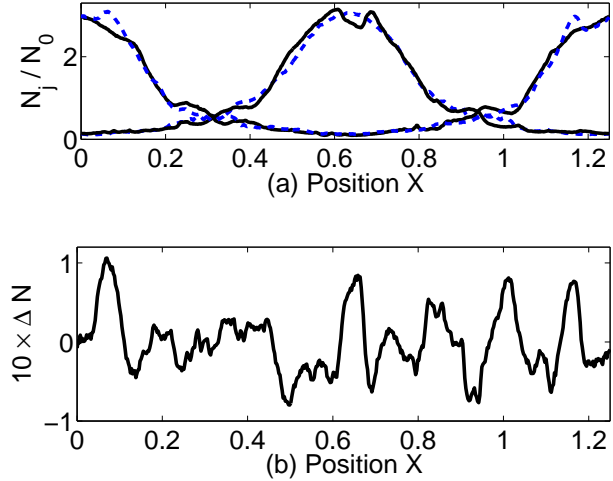}
\caption{(Colour online) Panel (a) The $N_j(x)$ at the time t=177 of the 
four species $j$ in units of $N_0$. The electron distributions $N_1(x)$ 
and $N_2(x)$ are denoted by the solid curves and $N_3, N_4$ are dashed
and blue. The curves $N_1 (x)$ and $N_4(x)$ peak at $x=0$. The curves for 
electrons and positrons do not match. Panel (b) plots the $\Delta N = 
\sum_j q_j N_j(x) / 4N_0$, with $q_j=1$ and $-1$ being the positron and 
electron charge. The fluctuation amplitude is about 10 percent and they 
oscillate on scales $\gg \Delta x$.}\label{Plot7}
\end{figure}

Figure \ref{Plot6} displays the phase space distributions of the species 
1, 2 and 4 at $t = 177$. The $f(x,v_z)$ are qualitatively unchanged 
compared to those in Fig. \ref{Plot4} and species 1 and 4 are still 
symmetric to a change of the sign of $v_z$. The distributions $f(x,v_x)$ 
reveal that the particles are concentrated at the same positions as in 
Fig. \ref{Plot4}. The particles have been heated up along the
x-direction in between the dense filaments. The small scale structures, 
in particular in $f(x,v_x)$, differ for species 1 and 4 and we expect
now clear charge density modulations.
Figure \ref{Plot7} compares $N_1 (x)$ with $N_4 (x)$ and $N_2(x)$ with 
$N_3(x)$. 
Their differences can now be seen even from the $N_i$. The value $N_4 
(x=0.08)$ exceeds that of $N_1 (x=0.08)$ by about 0.4 and fluctuations of 
this size occur in the entire box, which is demonstrated by the $\Delta N$
in Fig. \ref{Plot7}(b). The typical length scale of the oscillations is 
about $0.1 \gg \Delta x$ and they are caused by the waves in Fig. \ref{Plot3}. 

The peak amplitude of $E_x$ in Fig. \ref{Plot3} is $\approx 5 \times 
10^{-3}$ at $t=177$. The Lorentz force $v_b B_y$ for $v_b = 0.3$ and a 
maximum $B_y \approx 0.07$ at $t=177$ (Fig. \ref{Plot2}) is larger than 
the electrostatic force by a factor of four. The electrostatic force is
weaker before this time, while $B_y$ is constant after $t=50$. Thus, the 
electrons experience for most of the simulation duration the $E_x$ only 
as a perturbation and the gyrofrequency of the CPs is determined 
by $B_y$ and $v_z \approx v_b$, which is also demonstrated by the movie 1. 
The rotation of the dense filament in the distribution $f(x,v_x)$ has an 
approximately constant angular velocity in phase space. The structures in 
$f(x,v_z)$ and, thus, $J_z$ will change with this characteristic frequency 
and impose the modulation of $B_y$ and $E_z$ with $\Omega \approx 0.3$ in 
the Fig. \ref{Plot2}(d,f). This frequency is $\approx \Omega_i$, and 
consistent with the magnetic trapping.

The charge density fluctuations in Fig. \ref{Plot5}(b) at $t=50$ were at 
noise levels, which was expected from the low $E_{EX}$ in Fig. \ref{Plot1}
at this time and no QEI occured. It is only after the FI 
has saturated, that the electrostatic waves grow due to the SEI. The SEI 
yields the growth of $E_{EX}$ in the interval $50<t <177$ and the waves 
in Fig. \ref{Plot3} have no obvious correlation with the $B_y$ in Fig. 
\ref{Plot2}. We will now identify one potential cause of the SEI. The 
force imposed by a magnetic pressure gradient on a current $\mathbf{J}$ is 
\begin{equation}
\mathbf{J} \times \mathbf{B} = -\nabla \mathbf{B}^2 / 2.
\label{PrGr}
\end{equation}
Only $B_y$ is growing in our 1D simulation box and the only possible 
spatial derivative is $\partial_x$. Equation \ref{PrGr} simplifies to 
$J_z B_y = B_y dB_y / dx$. The $B_y$ is strong for $t>50$ and $dB_y / dx 
\neq 0$. The MPGF on the right-hand side does, therefore, not vanish 
and the particles are accelerated. 

\begin{figure}
\centering
\includegraphics[width=8.2cm]{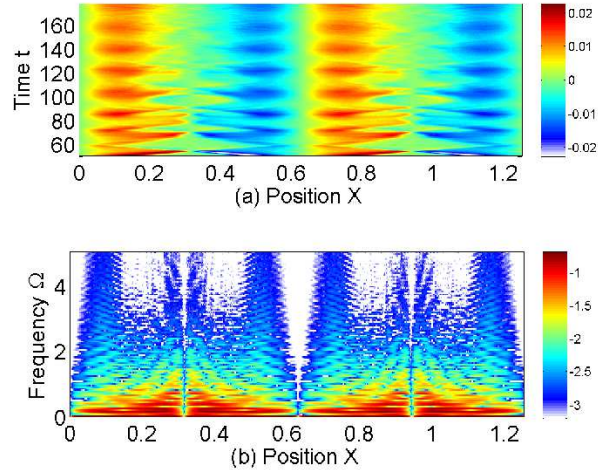}
\caption{(Colour online) Panel (a) shows the magnetic pressure gradient 
force $B_y dB_y / dx$, which oscillates twice as fast in space as $B_y$. 
The MPGF is not stationary in time. Panel (b) displays the base-10 
logarithm of the modulus of the spatial frequency spectrum $A(x,\Omega)$ 
of the MPGF. It is normalized to its peak value at 
$\Omega=0$. Strong oscillations close to $x=L/4$ and $x=3L/4$ reach 
$\Omega = 1$.}\label{Plot8}
\end{figure}

Figure \ref{Plot8} shows the MPGF after $t=50$ and $A(x,\Omega)$, which
is its spatial frequency spectrum obtained by a Fourier transform over time.
The maxima and minima of $\partial_x \tilde{P}_{BY}$ are stationary in 
space but its magnitude oscillates in time. The time average of the 
MPGF is positive to the right of the positions $x=0$ and $x=L/2$ and 
negative to the left of these positions. The $x=0$ and $x=L/2$ coincide 
with the stable equilibrium points of the respective filaments. The MPGF 
is thus accelerating the particles away from the equilibrium points. 
The MPGF is, however, weaker than the drift force $q_jv_bB_y$, which is 
responsible for the filament confinement (magnetic trapping). 

Consider the filament formed by the species 2 and 3 at $t=50$, which is 
centred at $x = L/2$ in Fig. \ref{Plot5} and Fig. \ref{Plot8}(a). The 
electrons of species 2 move with the velocity $\approx -v_b\mathbf{z}$, 
while the positrons of species 3 move with $\approx v_b \mathbf{z}$. 
Their currents have the same sign and the MPGF accelerates the electrons 
and the positrons into the same x-direction. As long as the positrons 
and the electrons have the same density everywhere, the $E_x$ does not
grow, because $J_x = 0$ in $\partial_t E_x + J_x = 0$. The term $\partial_y 
B_z -\partial_z B_y = 0$ in the 1D geometry. The MPGF does thus not result
here in the $E_x$-field discussed in Ref. \cite{Cal2,Dieckmann2}, which 
has a wavelength that is half of that of the wave in $B_y$. However, the 
finite number of CPs introduces statistical fluctuations in the charge 
density (Fig. \ref{Plot5}) that imply that the MPGF accelerates locally 
(on Debye length scales) a different number of electrons and positrons, 
by which $J_x \neq 0$. An electric field grows, that tries to restore the 
charge neutrality. The MPGF can couple to these fluctuations, if the force 
gradient is high (comparable to the spatial scale of the fluctuations) and 
if the force oscillates with the characteristic frequency of the fluctuations.
Their dominant oscillation frequency in the spatial intervals with $B_y 
\approx 0$ is the plasma frequency and it is the upper-hybrid frequency 
otherwise. Figure \ref{Plot3}(c) shows that the growing fluctuations have 
a broad frequency band and peak at $\Omega \approx 1$. The spatial gradients 
of the plasma density will, however, influence the fluctuation spectrum 
\cite{Belyi}. A force interacting with such fluctuations should thus
also have a broad frequency band. 

The spectrum $A(x,\Omega)$ in Fig. \ref{Plot8}(b) reveals that the 
highest frequencies can be reached by the MPGF at $x\approx 0.1,0.5,0.7$ 
and at $1.1$, although not with a high driving amplitude. The positions 
close to $x=L/4$ and $x=3L/4$ experience stronger oscillations up to 
$\Omega = 1$. The MPGF changes also on spatial scales comparable to the
Debye length, which is $v_t / c = 1/60$ in our normalization. An example 
is here a change at $t\approx 70$ and $x\approx 0.325$ in Fig. \ref{Plot8}(a). 
The MPGF changes from 0.02 to -0.02 over a distance $\approx 0.03$. The 
connection between the growth of $E_{EX}$ and the oscillations of the MPGF 
is evidenced also by a comparison of Fig. 
\ref{Plot1} with Fig. \ref{Plot8}(a). The growth of $E_{EX}$ slows down 
at $t \approx 100$. The oscillations in Fig. \ref{Plot8}(a) close to 
$L/4$ and $3L/4$ are more intense and shorter in duration before $t=100$ 
and the spatial gradients are higher. The MPGF can thus couple easier 
energy to the charge density fluctuations. The spatio-temporal 
oscillations of $\partial_x \tilde{P}_{BY}$ soften up after $t=100$, 
and the growth of $E_{EX}$ slows down. 

\section{Discussion}

In this paper we have considered the filamentation instability (FI) driven 
by two counterstreaming beams, each consisting of the electrons and positrons. 
The beams have initially been spatially uniform and the electromagnetic 
fields were set to zero. The 1D simulation is aligned with the x-direction,
which excludes the merging of filaments beyond a certain size and limits 
the physical realism of its nonlinear evolution \cite{Lee}. However, the 
filaments are usually not circular but elongated 
\cite{Dieckmann,Stockem,SilvaAIP}. A 1D geometry may thus be a valid 
approximation for those parts of filaments, which are quasi-planar such as 
the one investigated in Ref. \cite{Stockem}. The two beams move in opposite 
z-directions at the same speed modulus $v_b=0.3$, which is sufficiently low 
to exclude significant relativistic effects. It is sufficiently high to 
obtain a linear growth rate of the FI that is comparable to those of the 
electrostatic modes and of the mixed modes. This may be further aided
by the equal beam densities, which favor the filamentation instability
if positrons are absent \cite{Bellido}. The strong electric fields with 
their oblique polarization in the PIC simulation in Ref. \cite{Kazimura} 
demonstrate, however, that in particular the mixed modes would compete 
with the FI in a more realistic 2D or 3D simulation.

Our aim has been to obtain further insight into the dynamics of a filament 
pair formed by two counterpropagating beams of the electrons and positrons 
and to compare it with that of an electron filament pair. The simulation 
parameters are identical to those in the simulation in Ref. \cite{Dieckmann2} 
that did not take into account the positrons. The short simulation box allows 
us to use a good statistical plasma representation and low noise levels. The 
charge density fluctuations inherent to the PIC simulation method provide 
noise over the full band of wavenumbers resolved by the simulation, out of 
which the wavemodes and any secondary instabilities can grow.

The FI redistributes the beams of charged particles in space into current 
filaments \cite{Lee}. Our simulation box length $L$ and the periodic 
boundary conditions allow only a single pair of filaments to grow. The 
centres of the electron filaments are spatially separated by the distance 
$L/2$ and this is also the case for both positron filaments. The filament 
formed by the electrons of one beam coincides with the filament containing 
the positrons of the second beam. Their phase space distributions $f(x,v_x)$,
which hold the information about the electrostatic structures, match and
their phase space distributions $f(x,v_z)$ can be mapped from one to the
other by a change of the sign of $v_z$. The currents and densities of both 
components add up. The symmetry between the electrons and positrons within 
the same filament implies that the MPGF accelerates them into the same 
direction. We have found that their partial currents cancel each other
until the FI saturates, implying that no $E_x$ can grow due to the QEI. 

The simulation in Ref. \cite{Dieckmann2} evidenced that this $E_x$ would 
accelerate the electrons away from the centre of the filament. The presence 
of the positrons in the simulation in the present work removes this 
repulsion. Higher charge and current densities can be reached by 
electron-positron beams compared to those containing only electrons and 
the spacing between the filaments is larger. The magnetic fields can reach 
higher spatial gradients. The magnetic fields grew to an amplitude set
by magnetic trapping \cite{Davidson}.

We have identified a SEI that is leading to a growing electrostatic energy 
density after the FI saturated, resulting in a broadband wave spectrum. 
The exponential growth rates of these waves remained well below that of 
the FI and they show no correlation with the MPGF. These observations are 
conflicting with the properties attributed to the QEI, implying that the 
SEI and the QEI must have different source mechanisms.

The finite number of computational particles imply statistical fluctuations
of the charge density on Debye length scales. We have found here that the 
oscillation spectrum of the MPGF involves a wide band of frequencies that 
reach a maximum that exceeds the plasma frequency and that the MPGF 
changes significantly on scales comparable to the Debye length of the 
plasma after the FI has saturated. The MPGF can thus couple to the 
statistical fluctuations of the charge density. We have proposed that this 
coupling amplifies the fluctuations of $J_x$ and $E_x$ and that this is the 
cause of the SEI. 

This hypothesis can be tested and the properties of the SEI can be examined 
in more detail with Vlasov simulations. They 
solve the Vlasov-Maxwell equations directly and do not approximate the
plasma by phase space blocks. They are thus free of noise due to statistical 
plasma density fluctuations. 
The SEI should not develop in Vlasov simulations, unless seed fluctuations 
are introduced. The filaments, which are plasma structures resulting out 
of magnetic instabilities, would be more stable in Vlasov simulations than 
in PIC simulations. Previously, such an enhanced stability has only been 
reported for nonlinear structures (electron phase space holes) evolving 
out of purely electrostatic Buneman instabilities \cite{Comparison}. A 
localized charge density perturbation can be introduced into a Vlasov 
simulation and its interplay with the MPGF can be investigated. Such
studies can not be performed with PIC codes, where we always have an
ensemble of charge density perturbations interacting with the MPGF.
 
The energy density $E_{EX}$ of the waves driven by the SEI remained below 
$10^{-2} E_{BY}$. The $E_x$-field will, however, not be negligible after 
$t=177$, because the electric 
force $q_j E_x$ competes with the Lorentz force $q_j v_z B_y$ and $v_z < 1$. 
The ratio between the strongest Lorentz force and the strongest electric 
force has been 4 at $t=177$ and the instability is growing. 
The SEI is also likely to play a more important role in PIC simulations that 
use only a low number of particles per cell. The higher relative charge 
density fluctuations imply, that the SEI grows from higher initial amplitudes, 
which reduces its growth time. The initial electric noise power scales 
approximately inversely proportional to the number of particles per cell. 
A reduction from our $4.9 \times 10^4$ particles per cell to 50 particles 
per cell will increase the initial power of $E_x$ by the factor $10^3$.  

The energy density of the electric field has grown to a significant 
fraction of that of the magnetic field in the 1D and 2D simulations of 
counterpropagating electron beams \cite{Cal1,Cal2,Rowlands,Dieckmann2,Stockem}, 
and it should influence the interplay of the filaments during their nonlinear
evolution. No significant electric fields are observed during the quasi-linear 
growth phase of the FI and immediately after its saturation, if each beam
carries electrons and positrons with an equal density. Our future studies 
will thus assess the impact of an absent electric field on the filament size 
distribution. This distribution can be approximated by a Gumbel distribution 
in 1D if no positrons are present \cite{Rowlands}. We will also examine 
with 2D simulations how the filament size distribution orthogonal to the 
beam velocity vector evolves in time when positrons are present. The 
characteristic filaments size increases in the absence of the positrons 
linearly with time \cite{Dieckmann}. The impact of the electrostatic and 
mixed modes on the evolution of the filaments will also be addressed by 
2D simulation studies that contain the beam velocity vector. 

{\bf Acknowledgements} The authors would like to thank the research councils
of Sweden (Vetenskapsr\aa det) and Germany (DFG grant FOR1048) for financial 
support. The Swedish HPC2N computer centre has provided the computer time
and support.

\end{document}